\documentclass[twoside,twocolumn,9pt]{article}
\usepackage{extsizes}
\usepackage[super,sort&compress,comma]{natbib} 
\usepackage[version=3]{mhchem}
\usepackage[left=1.5cm, right=1.5cm, top=1.785cm, bottom=2.0cm]{geometry}
\usepackage{balance}
\usepackage{mathptmx}
\usepackage{sectsty}
\usepackage{graphicx} 
\usepackage{lastpage}
\usepackage[format=plain,justification=justified,singlelinecheck=false,font={stretch=1.125,small,sf},labelfont=bf,labelsep=space]{caption}
\usepackage{float}
\usepackage{fancyhdr}
\usepackage{fnpos}
\usepackage[english]{babel}
\addto{\captionsenglish}{%
  
}
\usepackage{array}
\usepackage{droidsans}
\usepackage{charter}
\usepackage[T1]{fontenc}
\usepackage[usenames,dvipsnames]{xcolor}
\usepackage{setspace}
\usepackage[compact]{titlesec}
\usepackage{hyperref}
\usepackage{upgreek}

\usepackage{color}
\usepackage{soul}

\usepackage{epstopdf}

\definecolor{cream}{RGB}{222,217,201}

\begin{document}

\pagestyle{fancy}
\thispagestyle{plain}
\fancypagestyle{plain}{
\renewcommand{\headrulewidth}{0pt}
}

\makeFNbottom
\makeatletter
\renewcommand\LARGE{\@setfontsize\LARGE{15pt}{17}}
\renewcommand\Large{\@setfontsize\Large{12pt}{14}}
\renewcommand\large{\@setfontsize\large{10pt}{12}}
\renewcommand\footnotesize{\@setfontsize\footnotesize{7pt}{10}}
\makeatother

\renewcommand{\thefootnote}{\fnsymbol{footnote}}
\renewcommand\footnoterule{\vspace*{1pt}%
\color{cream}\hrule width 3.5in height 0.4pt \color{black}\vspace*{5pt}} 
\setcounter{secnumdepth}{5}

\makeatletter 
\renewcommand\@biblabel[1]{#1}            
\renewcommand\@makefntext[1]%
{\noindent\makebox[0pt][r]{\@thefnmark\,}#1}
\makeatother 
\renewcommand{\figurename}{\small{Fig.}~}
\sectionfont{\sffamily\Large}
\subsectionfont{\normalsize}
\subsubsectionfont{\bf}
\setstretch{1.125} 
\setlength{\skip\footins}{0.8cm}
\setlength{\footnotesep}{0.25cm}
\setlength{\jot}{10pt}
\titlespacing*{\section}{0pt}{4pt}{4pt}
\titlespacing*{\subsection}{0pt}{15pt}{1pt}

\fancyfoot{}
\fancyfoot[LO,RE]{\vspace{-7.1pt}\includegraphics[height=9pt]{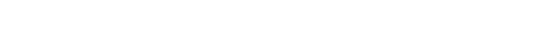}}
\fancyfoot[CO]{\vspace{-7.1pt}\hspace{13.2cm}\includegraphics{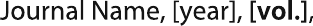}}
\fancyfoot[CE]{\vspace{-7.2pt}\hspace{-14.2cm}\includegraphics{head_foot/RF}}
\fancyfoot[RO]{\footnotesize{\sffamily{1--\pageref{LastPage} ~\textbar  \hspace{2pt}\thepage}}}
\fancyfoot[LE]{\footnotesize{\sffamily{\thepage~\textbar\hspace{3.45cm} 1--\pageref{LastPage}}}}
\fancyhead{}
\renewcommand{\headrulewidth}{0pt} 
\renewcommand{\footrulewidth}{0pt}
\setlength{\arrayrulewidth}{1pt}
\setlength{\columnsep}{6.5mm}
\setlength\bibsep{1pt}

\makeatletter 
\newlength{\figrulesep} 
\setlength{\figrulesep}{0.5\textfloatsep} 

\newcommand{\topfigrule}{\vspace*{-1pt}%
\noindent{\color{cream}\rule[-\figrulesep]{\columnwidth}{1.5pt}} }

\newcommand{\botfigrule}{\vspace*{-2pt}%
\noindent{\color{cream}\rule[\figrulesep]{\columnwidth}{1.5pt}} }

\newcommand{\dblfigrule}{\vspace*{-1pt}%
\noindent{\color{cream}\rule[-\figrulesep]{\textwidth}{1.5pt}} }

\makeatother

\twocolumn[
  \begin{@twocolumnfalse}
{\includegraphics[height=30pt]{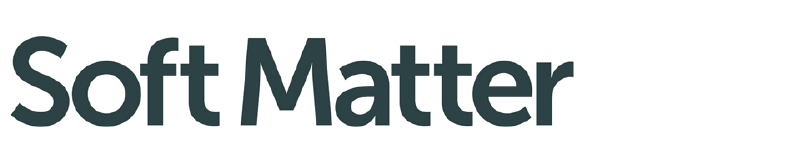}\hfill\raisebox{0pt}[0pt][0pt]{\includegraphics[height=55pt]{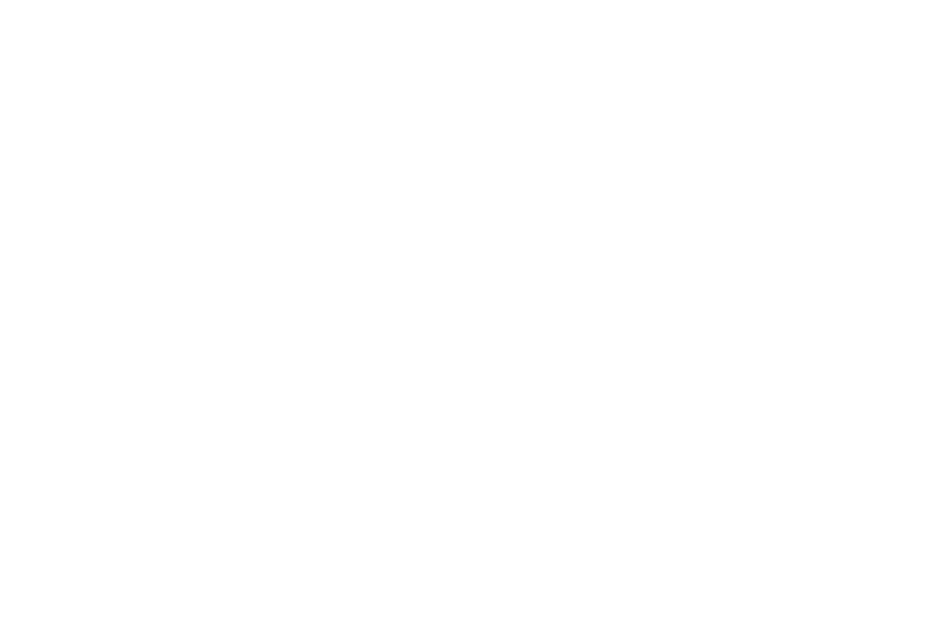}}\\[1ex]
\includegraphics[width=18.5cm]{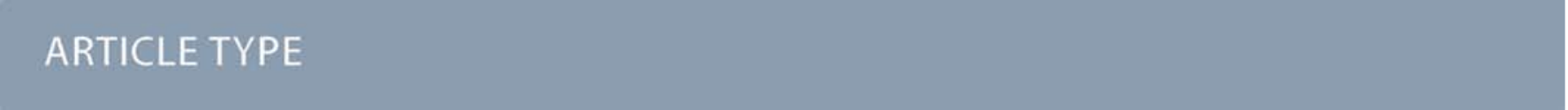}}\par
\vspace{1em}
\sffamily
\begin{tabular}{m{4.5cm} p{13.5cm} }

\includegraphics{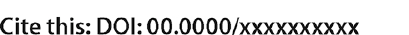} & \noindent\LARGE{\textbf{Bouncing-to-wetting transition for impact of water droplets on soft solids}} \\
\vspace{0.3cm} & \vspace{0.3cm} \\

 & \noindent\large{Surjyasish Mitra \textit{$^{a}$}, Quoc Vo \textit{$^{{b},{\ast}}$}, 
 and Tuan Tran \textit{$^{b,{\ast \ast}}$}} \\

\includegraphics{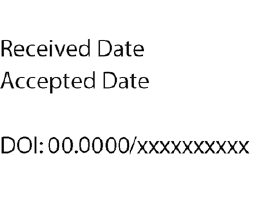} & \noindent\normalsize
{Soft surfaces impacted by liquid droplets
trap more air underneath 
than their rigid counterpart. 
The extended lifetime 
of the air film 
not only facilitates bouncing behaviours 
of the impacting droplets
but also increases the
possibility of 
an interaction between the 
air film itself and 
the air cavity formed 
inside the droplets 
by capillary waves.
Such interaction
may cause rupture of 
the trapped air film
by a so-called dimple inversion phenomenon
and suppress bouncing.
In this work, 
we systematically investigate 
the relation between 
air cavity collapse and 
air film rupture for 
water droplets impacting on
soft, hydrophobic surfaces.
By constructing a bouncing-to-wetting phase diagram 
based on the rupturing dynamics 
of the trapped air film, 
we observe that
the regime in which 
air film rupture is induced 
by dimple inversion
consistently separates the
bouncing regime and the 
one in which wetting is caused by
random rupture.
We also find that
air film rupture 
by dimple inversion, in-turn,
affects both the collapsing dynamics 
of the air cavity and 
the resulting high-speed jet.
We then provide a detailed characterisation 
of the collapsing dynamics 
of the air cavity 
and subsequent jetting.}  \\

\end{tabular}

 \end{@twocolumnfalse} \vspace{0.6cm}

  ]

\renewcommand*\rmdefault{bch}\normalfont\upshape
\rmfamily
\section*{}
\vspace{-1cm}


\footnotetext{\textit{$^{a}$ School of Physical \& Mathematical Sciences, Nanyang Technological University, 50 Nanyang Avenue, 639798 Singapore}}
\footnotetext{\textit{$^{b}$ School of Mechanical \& Aerospace Engineering,  Nanyang Technological University, 50 Nanyang Avenue, 639798 Singapore. 
E-mail:$^{\ast}$ xqvo@ntu.edu.sg, $^{\ast \ast}$ ttran@ntu.edu.sg }}


\section{Introduction}
The air film separating an impacting liquid droplet
with an impacted surface plays a crucial role
in dictating impact outcomes. 
For high velocity impacts,
the lubrication pressure built-up in the air film
causes liquid splashes \cite{xu2005drop,driscoll2011ultrafast}.
For low velocity impacts, 
the presence of a
sustained air film leads 
to bouncing of the droplets 
regardless of 
surface's wettability \cite{kolinski2012skating,de2012dynamics,
de2015wettability,de2015air,van2012direct}.
Studies of the bouncing-to-wetting transition
of low velocity water droplets impacting 
on solid surfaces 
are motivated by both fundamental 
and practical interests. 
The former motivation comes from 
numerous intangible physical phenomena, 
e.g., capillary waves \cite{yarin2006drop}, 
air entrapment dynamics  \cite{de2015wettability},
involved in dictating
such transition, 
while the latter 
one is from applications requiring 
design of dynamic surfaces 
such as anti-bacterial, 
self-cleaning surfaces \cite{bhushan2011natural} 
or improvement of industrial processes 
including droplet deposition \cite{breitenbach2018drop} 
and ink-jet printing
\cite{ashgriz2011handbook}.
Although numerous investigations have
been focusing on the bouncing-to-wetting transition, 
in particular on the dynamics of
the intervening air film 
at the moment wetting occurs 
\cite{kolinski2012skating,
de2012dynamics,de2015wettability,
de2015air,van2012direct,li2015time},
the mechanisms
causing such transition 
remains elusive. 

Typically, the prelude 
to the final touchdown between 
an approaching droplet and a solid surface
is the formation of a thin air film 
in which lubrication pressure is built up. 
The lubrication pressure subsequently 
becomes sufficiently large that it 
deforms the droplet's bottom surface,
creating a central dimple surrounded by  
an outer edge with one or two kinks, 
the regions where the air film thickness is minimum \cite{klaseboer2014universal,
de2015air,de2015wettability}. 
For impact on hydrophilic surfaces,
bouncing is ensured 
at low impact velocity.
For higher impact velocity,
the air film typically ruptures 
either at inner or outer kink 
leading to wetting initiation \cite{de2015air,de2015wettability}.
In other words, bouncing-to-wetting transition for droplets impacting 
smooth hydrophilic surfaces, e.g.,  
glass or mica, 
is mainly determined by 
air film's dynamics 
and the surface properties
that cause random wetting initiation \cite{de2015air,de2015wettability}.

\begin{figure}[t]
  \begin{center}
    \includegraphics[width=.45\textwidth]{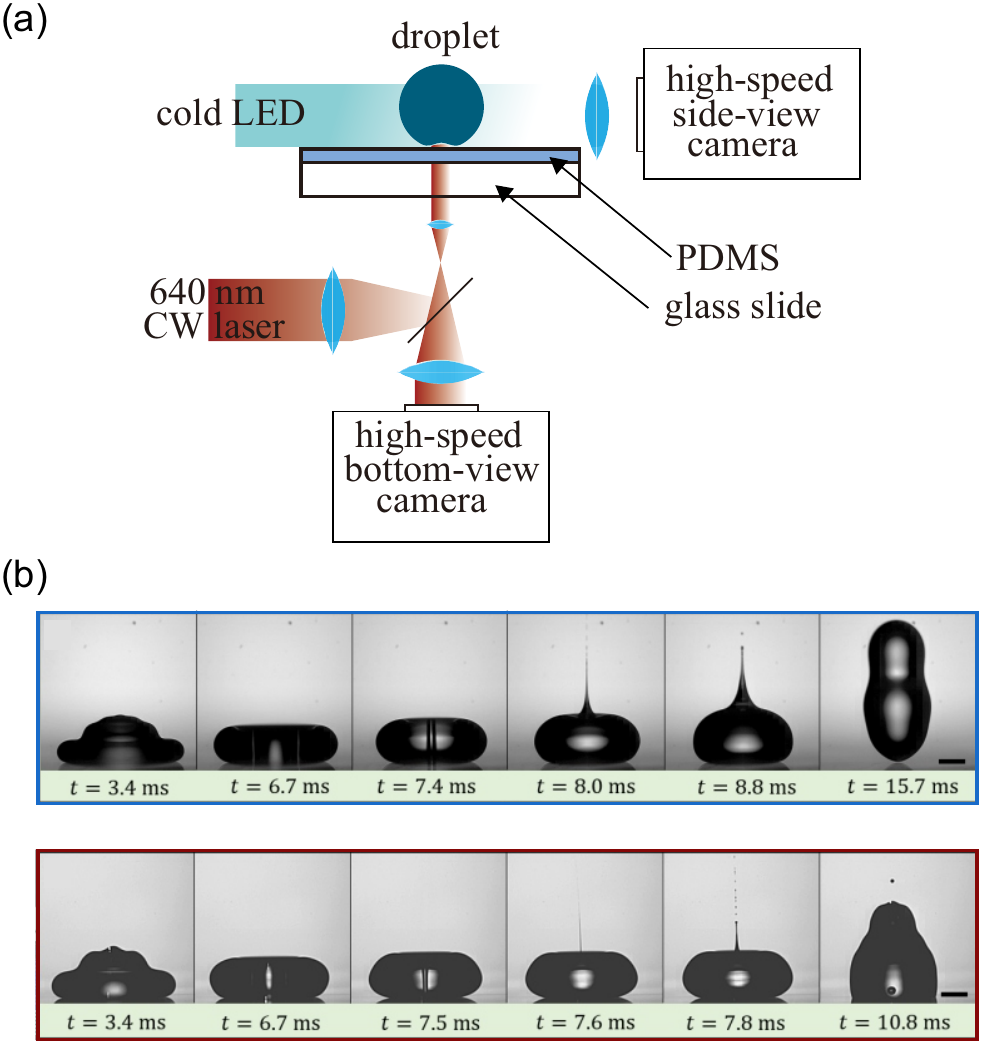}
    \end{center}
\caption{(a) Schematic of the experimental setup.
(b) Impact dynamics of a 1.16\,mm 
water droplet on soft PDMS 60:1 
substrates with Young's modulus $E = 4.8\,{\rm kPa}$ and  
two impact velocities: 
0.33 ${\rm m\, s^{-1}}$ ($\rm{We} = 1.7$) (upper panel), 
0.39 ${\rm m\, s^{-1}}$ ($\rm{We} = 2.4$) (lower panel). 
The snapshots corresponding to the last column
in both panels 
show two different impact outcomes:
bouncing and deposition with entrapped air bubble.
Scale bars represent 0.5\,mm.}	
\label{fig:1}
\end{figure} 

Impact outcomes 
on soft surfaces
are markedly different 
\cite{chen2011evolution,
chen2016droplet,langley2020droplet}.
On the one hand, an elastomer or gel coated surface
is usually hydrophobic \cite{andreotti2020statics}.
Thus, impacts of water droplets
on such soft surfaces appear to 
have characteristics similar to 
those on smooth, rigid hydrophobic surfaces \cite{bartolo2006singular,chen2017submillimeter}. 
An impacting droplet 
above a critical impact velocity 
may develop a pyramidal structure 
due to capillary waves induced upon impact. 
This leads to formation of 
a cylindrical air cavity that 
penetrates deeply into the droplet 
and eventually collapses, 
shooting out a liquid jet 
\cite{chen2011evolution,
chen2016droplet,bartolo2006singular,chen2017submillimeter}.
On the other hand, 
due to the deformation 
of the soft solids, 
air film trapped beneath 
the impacting droplets
have a longer lifetime compared 
to those impacting on a rigid one \cite{langley2020droplet}.
The sustained air film not only
inhibits wetting initiation
and facilitates 
bouncing of the impacting droplet \cite{chen2011evolution,
chen2016droplet},
but also 
increases the probability 
for interaction between 
the air film 
and the air cavity formed 
in the liquid bulk.
We note that the latter condition 
is rarely met 
for impacts on smooth, rigid hydrophilic
or hydrophobic surfaces due to
random air film rupture and wetting initiation \cite{de2015air,de2015wettability,
bartolo2006singular,chen2017submillimeter}.
Whether the interactions
between the trapped air film 
and the air cavity
dictates the wetting initiating mechanism
of the bouncing-to-wetting 
transition for impact of 
water droplets 
on soft surfaces 
remains an open question.


In this work, 
we experimentally study 
the bouncing-to-wetting transition of
water droplets impacting on soft surfaces.
We hypothesise that
such 
transition is determined 
by interactions between 
the air cavity and the air film 
trapped beneath the impacting droplets.
To test this hypothesis,
we first use both high-speed laser 
interferometry 
and high-speed optical imaging to
characterize 
the rupture mechanism 
of the air film and the dynamics 
of the air cavity in the bulk.
We reveal how 
interaction between 
the air film 
and the cavity's collapsing dynamics 
results in different types of 
air film rupture.
We then construct 
a phase diagram describing 
the dependence of 
the bouncing-to-wetting transition 
on the stiffness of the 
substrate 
and the impact characteristics 
based on the characterized 
air film rupture mechanisms.
Finally, we provide an explanation 
for the collapsing dynamics 
of the air cavity
and the resulting 
high-speed jets and
bubble entrapment.

\vskip -0.5cm
\section{Experiments}
Our test substrates
were glass slides coated
with a layer of Polydimethylsiloxane (PDMS)
of thickness $70\,\upmu$m.
The elasticity $E$ of the substrate 
was varied from $4.8\,$kPa to $263.6\,$kPa
by changing the ratio
of the monomer to crosslinker from 60:1 to 30:1 (Tab.~\ref{tb:material}).
The equilibrium contact angle $\theta_{\rm Y}$
of a small water droplet
on the test substrates
varied between 103.6$^{\circ}$ and 113.0$^{\circ}$ (Tab.~\ref{tb:material}).
Impact experiments were conducted 
using water droplets 
with radius $r_{0}$ varying from $0.63\,$mm 
to $1.4\,$mm; the velocity 
$v$ of the impacting droplets was varied between 
$0.30$\,${\rm m\, s^{-1}}$ 
and $0.51$\,${\rm m\, s^{-1}}$.  
The Weber number, defined as
${\rm We} = \rho r_{0} v^2/\gamma$,
thus varies 
from $2.1$ to $5.1$.
Here, $\rho = 1000\,{\rm kg\,m^{-3}}$
and $\gamma = 72\, {\rm N m^{-1}}$
respectively are the density 
and surface tension of water.

\begin{table*}[h]
\caption {Substrates used in our experiments.
The numbers show the weight ratio of the monomer to crosslinker. The value of stiffness $E$ is measured experimentally using a rheometer.
Equilibrium contact angles are measured using the sessile drop technique.}
 \label{tb:material}
\begin{center}
 \begin{tabular}{c c c c c c} 
\hline\hline 
 Substrates: & P60:1 & P50:1 & P40:1 & P35:1 & P30:1\\[0.5ex]
 \hline 
 $E$ (kPa) & 4.8 & 29.5 & 70.9 & 142.7 & 263.6\\[0.5ex]
$\theta_{\rm Y}$ & 109.5$^{\circ}$\,$\pm$\,3.5$^{\circ}$ & 108.2$^{\circ}$\,$\pm$\,1.7$^{\circ}$ & 107.7$^{\circ}$\,$\pm$\,0.8$^{\circ}$ & 107.0$^{\circ}$\,$\pm$\,1.3$^{\circ}$ & 106.8$^{\circ}$\,$\pm$\,3.2$^{\circ}$ \\[0.5ex]
 \hline
\end{tabular}
\end{center}
\end{table*}

Impacting droplets were recorded 
synchronously 
from the bottom and the side 
using two high-speed cameras 
(SA-X2 and SA-5, Photron) 
operating at imaging rates 
from 30,000 to 
200,000 frames-per-second 
and shutter time 1/800,000. 
This setup has 
a temporal resolution 
at 3.75\,$\mu$s. 
A cold LED light source was 
used for side view illumination, 
while a red diode laser 
(wavelength $\lambda = 640$\,nm) 
was used to illuminate 
the impacted surface from below (Fig.~\ref{fig:1}a). 
The laser illumination coupled with a 
5$\times$ optical zoom lens 
enabled us to observe the 
dimple profile with a 
height resolution of $\lambda/4 = 160$\,nm 
\cite{de2015air,hendrix2012spatiotemporal}, 
and a horizontal resolution of 3.5 $\mu$m/pixel.

\section{Results and Discussions}
\subsection{Bouncing-to-wetting transition}
For impacts of low viscosity liquids, 
the intrinsic length scale 
of the capillary waves induced upon impact 
is the wavelength 
${\rm{\lambda}_c} \sim \gamma{\rho}^{-1}{v}^{-2}$ 
\cite{bartolo2006singular}.
The condition for capillary wave formation 
is $\lambda_{\rm c} > r_0$, or equivalently 
${\rm We } > 1$. 
As the Weber number in our experiment varies 
between 2 and 5,  
both inertial and 
capillary forces are significant. 
In this Weber number range, 
impacting droplets deform 
into pyramidal shapes, as shown in  
Fig.~\ref{fig:1}b ($t = 3.4\,$ms).

From side-view images, we observe 
two distinct macroscopic behaviours
when the velocity
$v$ of the impacting droplet increases: 
the droplet either bounces
off from the substrate at low impact velocity (Fig.~\ref{fig:1}b, upper panel) 
or is deposited onto the substrate as the
impact velocity in increased (Fig.~\ref{fig:1}b, lower panel).
In both cases, 
we observe 
that the air cavity forms
roughly at the moment
the droplet reaches
the maximum deformation and 
starts retracting 
(Fig.~\ref{fig:1}b, $t = 6.7\,$ms).
Subsequently, the air cavity 
collapses resulting in 
liquid jets from the droplet.
Typically, we observe that the jet velocity
is higher in the case of depositing droplets 
compared to that of bouncing droplets from the substrate. 

\begin{figure*}[!h]
  \begin{center}
    \includegraphics[width=0.9\textwidth]{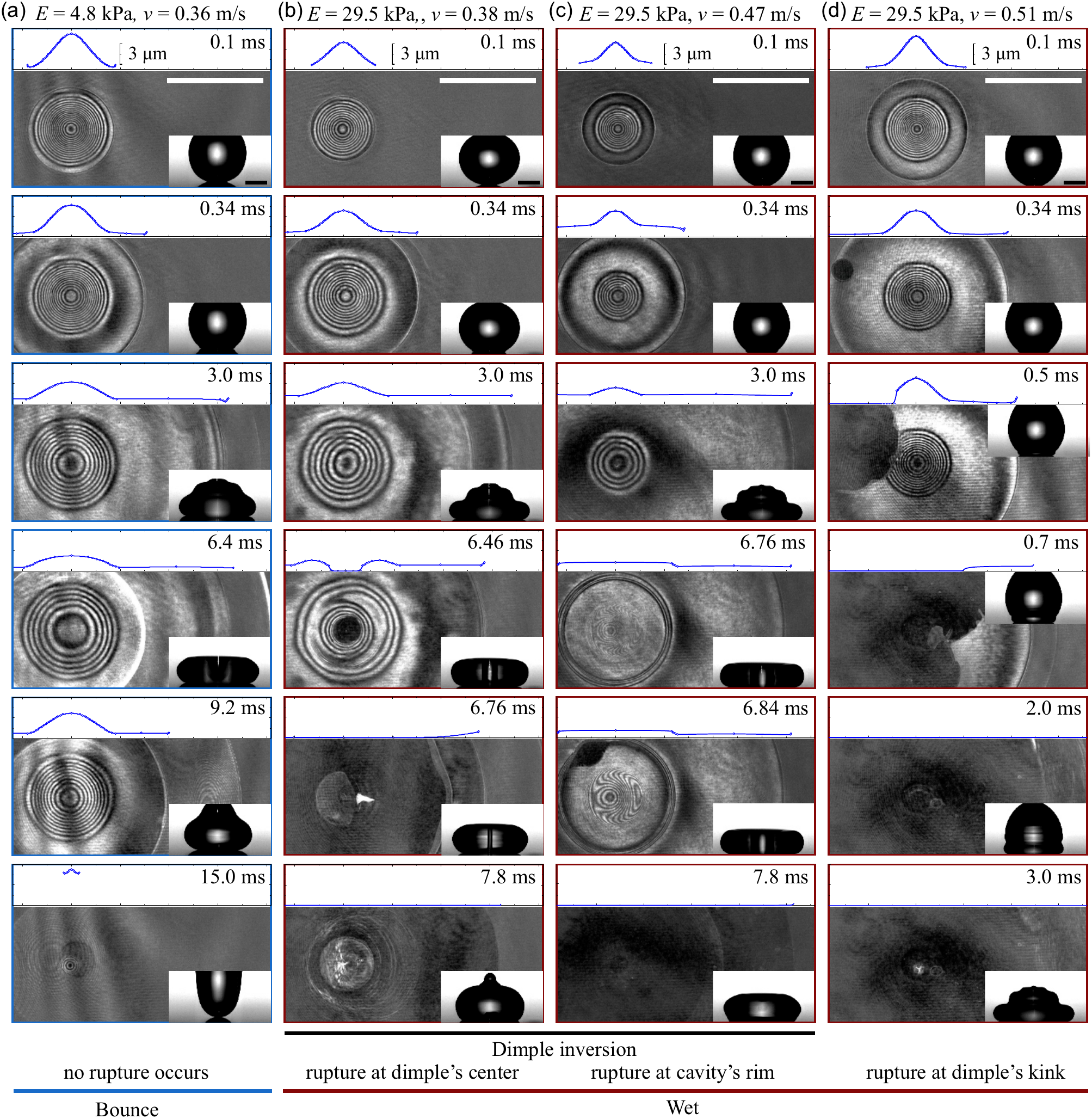}
    \end{center}
\caption{
Air film rupture dynamics of 1.4\,mm 
radius water drops impacting on soft PDMS surfaces 
at different impact velocity $v$.
(a) At $v = 0.36\,{\rm m\, s^{-1}}$, 
the droplet bounces of after 
impact without wetting the solid surface. 
(b) At $v = 0.38\,{\rm m\, s^{-1}}$, 
a perfect inversion of the air cavity causes 
rupture of the air film
at the dimple's center.
Liquid jet and air bubble entrapment 
caused by the collapse of air cavity are observed.
(c) At $v = 0.47\,{\rm m\, s^{-1}}$,
rupture of air film occurs at the rim of the air cavity. 
(d) At $v = 0.51\,{\rm m\, s^{-1}}$,
rupture of air film occurs randomly 
at the rim location
where the air film is thinnest. 
The scale bars in both side-view and 
bottom-view images represent 1\,mm. 
}
\label{fig:2}
\end{figure*}

To reveal mechanisms causing the 
different impact outcomes, i.e., bouncing or deposition, 
in Fig.~\ref{fig:2}, we show 
bottom-view interferometric recordings 
and the corresponding extracted profiles
of the air film trapped between 
the liquid and the solid upon impact. 
In all cases, 
the liquid droplet initially skates 
on a thin film of air 
with the bottom liquid-air interface 
expanding at a lateral speed 
$\sim$\,0.2\,${\rm m\, s^{-1}}$, 
in close proximity to those reported 
in a recent study 
for impacts in a similar range of Weber number \cite{pack2019role}. 
While expanding, 
the bottom surface of the droplet 
deforms into the familiar dimple profile 
\cite{de2012dynamics,de2015air} 
with two distinct kinks 
due to excess pressure in the trapped air film, 
evident from the recorded 
interference signatures (see Fig.~\ref{fig:2}\,a-d, $t = 0.34$\,ms).

The evolution of the trapped air film
is sensitive to the impact velocity.
At low impact velocity, 
$v \le 0.36\,{\rm m\,s^{-1}}$ 
(Fig.~\ref{fig:2}a), 
the air film remains intact 
during the entire duration of impact.
Therefore, 
the droplet bounces off 
from the substrate similar 
to previous studies involving
droplet bouncing
on glass \cite{de2015wettability}
and PDMS \cite{chen2016droplet} surfaces.
For $0.38\,{\rm m\,s^{-1}} \le v \le 0.47\,{\rm m\,s^{-1}}$, 
the air cavity in the bulk
forces the dimple downward 
causing shape inversion.
The air film 
separating the inverted 
dimple and the soft solid 
eventually ruptures, 
initiating wetting.
We observe 
two distinct types of
wetting initiation
caused by the inversion of the dimple:
wetting either first occurs at the 
center of the dimple 
(Fig.~\ref{fig:2}b, $t = 6.76$\,ms),
or at the dimple's inner rim 
(Fig.~\ref{fig:2}c, $t = 6.84$\,ms).
While wetting initiation 
at dimple's center 
had been previously observed 
in a similar study of droplet impact on PDMS surfaces \cite{chen2011evolution},
wetting initiation at dimple's inner rim due to dimple inversion has not been reported.
In our experiment, the wetting initiation 
at the dimple's inner rim
consistently happens at impact
velocity slightly higher 
than that at the dimple's center.
For a higher range of impact velocity, 
$v \geq 0.51\,{\rm m\,s^{-1}}$, 
rupture of the air film initiates 
at a random position near
the inner or outer kink 
(Fig.~\ref{fig:2}d, $t = 0.34$\,ms)
consistent with previous
drop impact studies on rigid
and soft surfaces \cite{de2012dynamics,de2015air,chen2016droplet}. 

\begin{figure}[t]
  \begin{center}
    \includegraphics[width=0.45\textwidth]{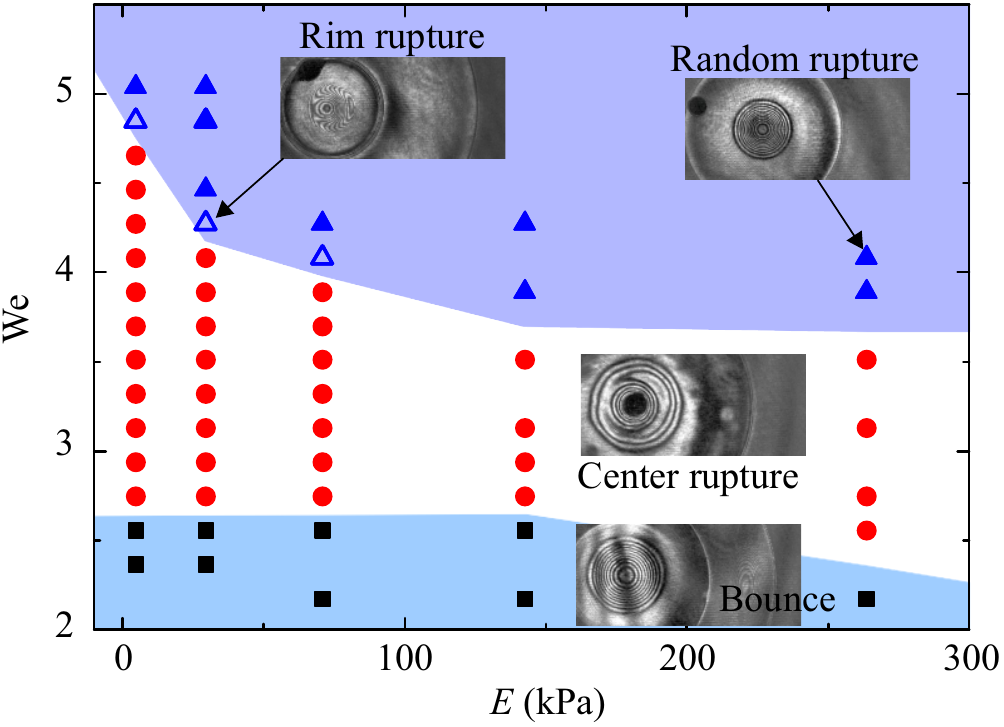}
          \end{center}
\caption{ Phase diagram showing 
the general behaviours, i.e., bouncing, dimple inversion and wetting, of droplets impacting on soft solid. 
The behaviours are obtained by varying Weber number ${\rm We}$ and Young modulus $E$ of the soft substrates. 
The droplet is water having fixed radii $r_{0} = 1.4\,$mm.}	
\label{fig:3}
\end{figure}

\begin{figure*}[t]
  \begin{center}
    \includegraphics[width=0.92\textwidth]{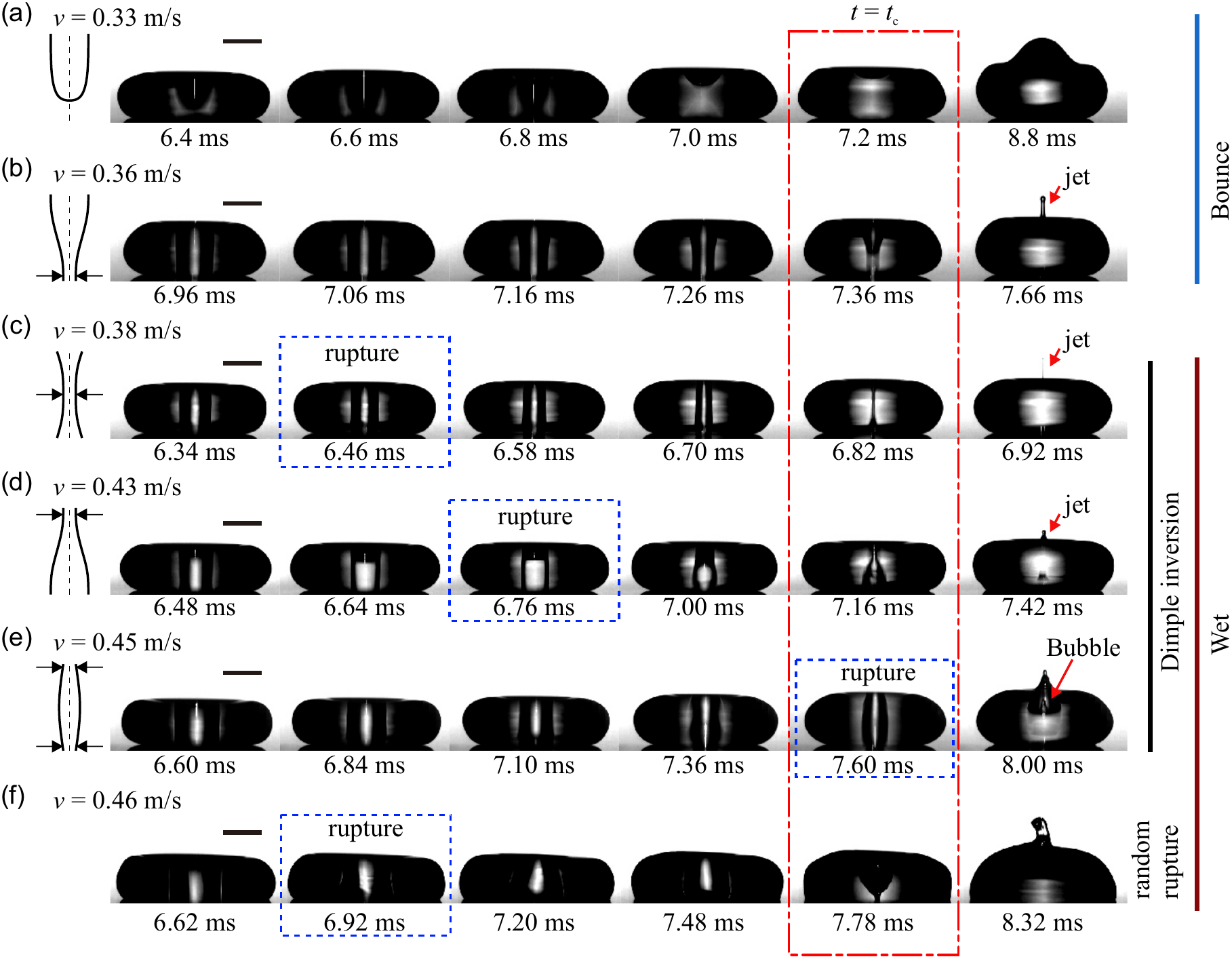}
    \end{center}
\caption{Snapshots showing the formation
and collapse of air cavity during 
impacts of  
$r_{0} = 1.4\,$mm water droplets onto PDMS 40:1 substrates with $E = 70.9\,{\rm kPa}$ at different impact velocities $v$.
From (a) to (f), 
the impact velocity $v$ increases
from $0.33\,{\rm m\, s^{-1}}$ to $0.46\,{\rm m\, s^{-1}}$ 
resulting in different collapse dynamics of the air cavity.
The dashed boxes show 
the snapshots at the moment 
the air film ruptures.
The schematics show 
the air cavity shapes right before collapse, i.e., $t = t_{\rm c}$ 
(see the dashed-dotted box).
The arrows in each schematic 
show the position 
where the collapse occurs.
In all cases, the scale bars represent 1\,mm.}
\label{fig:4}
\end{figure*}

In Fig.~\ref{fig:3}, 
we show a phase diagram
presenting how the 
Weber number ${\rm We}$
and surface elasticity $E$ 
affect the behaviour 
of the trapped air film 
and the resulting impact outcomes.
Generally, we observe 
that dimple inversion
causing air film rupture
at the center of the dimple 
is sandwiched between two other major behaviours:
bouncing and random rupture. 
Wetting initiated by rupture at the 
dimple's rim (open triangles) appears as
a transitional behaviour separating the  
center rupture and the random rupture behaviours.
In other words,
for a fixed soft substrate, increasing the
${\rm We}$ number
causes the impact dynamics
to change
from bouncing to center rupture, rim rupture and finally 
random rupture.
For surfaces with $E > 142$\,kPa, 
we do not observe
the rim rupture behaviour. 
We also observe that 
increasing the substrate elasticity $E$, 
while causing insignificant effect to the 
transition between bouncing and wetting
considerably reduces the 
transitional ${\rm We}$ number
between center rupture and rim rupture.
This is qualitatively 
consistent with the fact that increasing the surface stiffness
results in less air entrapment and thinner air film, 
which eventually leads 
to random rupture of the air film
and wetting initiation\cite{de2015wettability,de2015air}. 

\subsection{Collapse dynamics of air cavity}

The dimple and 
air film dynamics
not only affect the 
general impact outcomes 
but also alter
the formation and collapse 
of the air cavity formed
upon impact.
In Fig.~\ref{fig:4},
we show several series of snapshots
highlighting the formation 
and collapse of the 
air cavity 
upon impact of 
$r_{0} = 1.4\,$mm water droplets 
on soft PDMS surfaces having $E = 70.9\,\rm{kPa}$.
At low impact velocity, 
i.e., $v = 0.33\,{\rm m\,s^{-1}}$,
we observe that 
a U-shaped cavity is partially formed (Fig.~\ref{fig:4}a).
Subsequently, the droplet's surface 
restores to its minimum surface area 
pushing the cavity out.
No liquid jet or 
air bubble entrapment
inside the liquid bulk is observed.
At $v \ge 0.36\,{\rm m\,s^{-1}}$,
the capillary waves on 
the droplet's surface
are sufficiently strong 
to form a cylindrical cavity 
through the droplet width 
(Fig.~\ref{fig:4}b-f, the first snapshots).
Subsequently the cavity radius retracts and eventually collapses 
resulting in liquid jet and bubble entrapment.
Depending on the impact velocity $v$,
the cavity either collapses at
the bottom of 
the cylindrical cavity (Fig.~\ref{fig:4}b, $t = 7.36\,{\rm ms}$), 
or at the cavity's waist 
(Fig.~\ref{fig:4}c, $t = 6.82\,{\rm ms}$), 
cavity's top (Fig.~\ref{fig:4}d, $t = 7.16\,{\rm ms}$)
or at both top and bottom (Fig.~\ref{fig:4}e, $t = 7.60\,{\rm ms}$).
We attribute the different collapsing dynamics
of the cavity to 
the capillary waves 
on the cavity surface 
which is
clearly observed in 
(Fig.~\ref{fig:4}b-e):
the collapse position 
occurs at the wave's peak.
At high impact velocity, 
where the air film ruptures randomly,
(Fig.~\ref{fig:4}f),
the wetting initiation
at the rupture point 
disturbs the retracting cavity
in the bulk.

\begin{figure}[t]
  \begin{center}
    \includegraphics[width=0.37\textwidth]{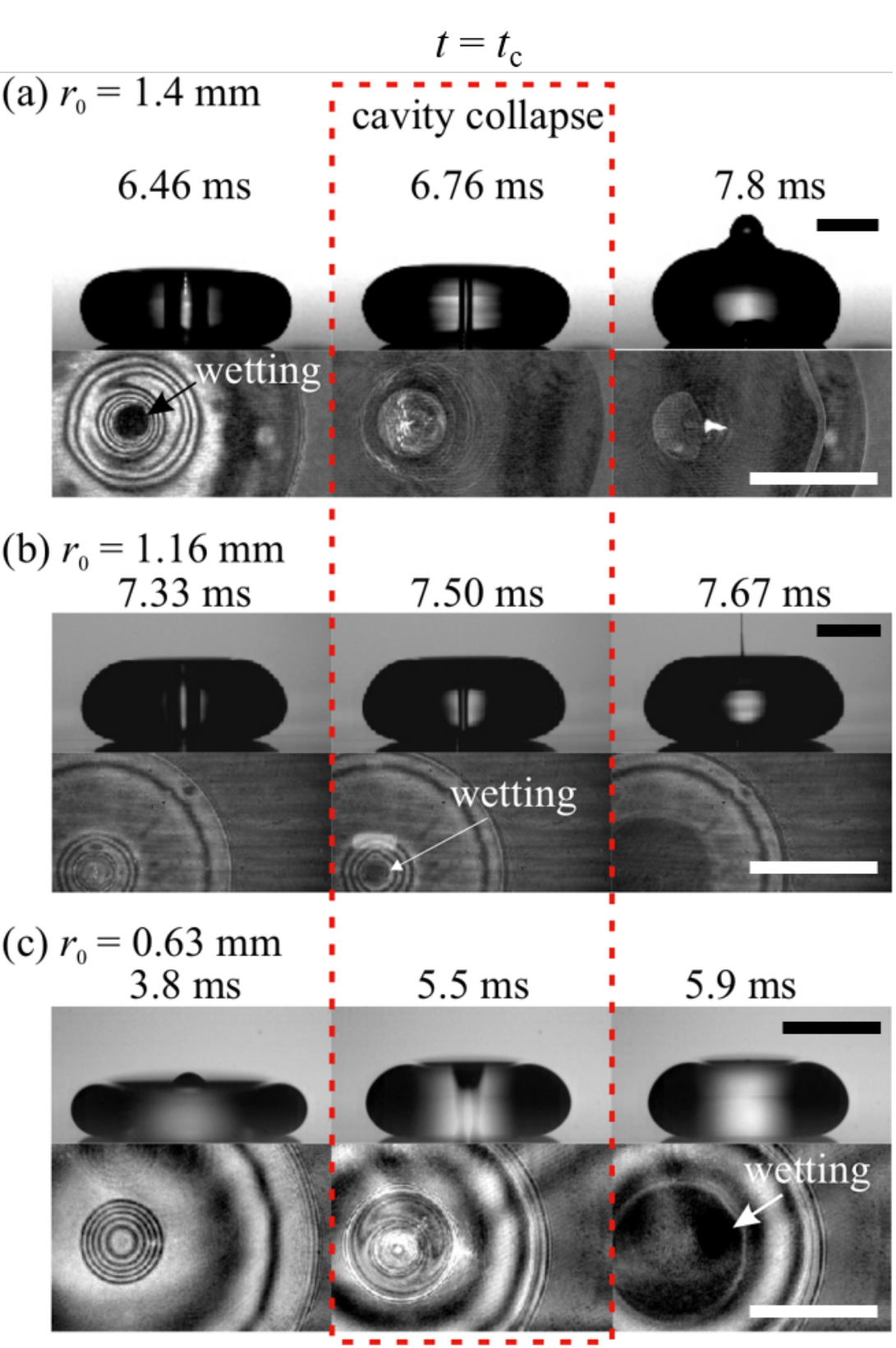}
          \end{center}
\caption{Snapshots showing 
changes in temporal order 
of air cavity collapse 
and wetting initiation
by varying droplet radius: 
(a) $r_0 = 1.4\,$mm, wetting occurs before cavity collapse; 
(b) $r_0 = 1.16\,$mm, wetting initiates right at the moment cavity collapses; 
(c) $r_0 = 0.63\,$mm,  wetting initiates after cavity completely collapsed. 
Parameters: (a) $E = 29.5\,$kPa,
 ${\rm We} = 2.8$; (b) $E = 4.8\,$kPa, ${\rm We} = 2.4$; (c) $E = 4.8\,$kPa, ${\rm We} = 2.3$.
Scale bars
in both the side-view and 
bottom-view snapshots represent 
1\,mm.}
\label{fig:3New}
\end{figure}

Interestingly, 
we observe that the 
relative 
temporal order between wetting initiation 
and complete collapse of the air cavity 
depends on both impact velocity and droplet radius.
As shown in Fig.~\ref{fig:4}c-e,
with a fixed droplet radius ($r_0 = 1.4\,$mm),
the duration $t_{\rm c} - t_{\rm r}$ 
decreases from $0.36\,$ms to $0$\,ms 
when the impact velocity $v$ 
increases from $0.38\,{\rm m\,s^{-1}}$ (Fig.~\ref{fig:4}c) to $0.45\,{\rm m\,s^{-1}}$ (Fig.~\ref{fig:4}e).
Here, $t_{\rm r}$ is the moment at which
rupture by dimple inversion happens (dashed-blue boxes in Fig.~\ref{fig:4}).
In Fig.~\ref{fig:3New}, we show that 
wetting occurs \emph{before} 
the final collapse of the air cavity for 
larger droplets (see Fig.~\ref{fig:3New}a,
$r_{0} = 1.4$\,mm) and \emph{after} the collapse
of the air cavity for smaller droplets
(see Fig.~\ref{fig:3New}c,
$r_{0} = 0.63$\,mm).
The radius at which
the temporal order between wetting initiation 
and cavity collapse switches 
is 
$r_{0} = 1.16$\,mm (Fig.~\ref{fig:3New}b).
This is consistent with 
a previously reported experimental observation by 
Chen et~al. \cite{chen2011evolution} 
in which wetting initiation 
at the dimple center 
occurs after 
the air cavity completely collapses
for water droplets of radius $r_{0} = 1\,$mm and a similar Weber number
impacting 
on PDMS coated substrates.
%

\begin{figure*}[t]
\begin{center}
\includegraphics[width=0.9\textwidth]{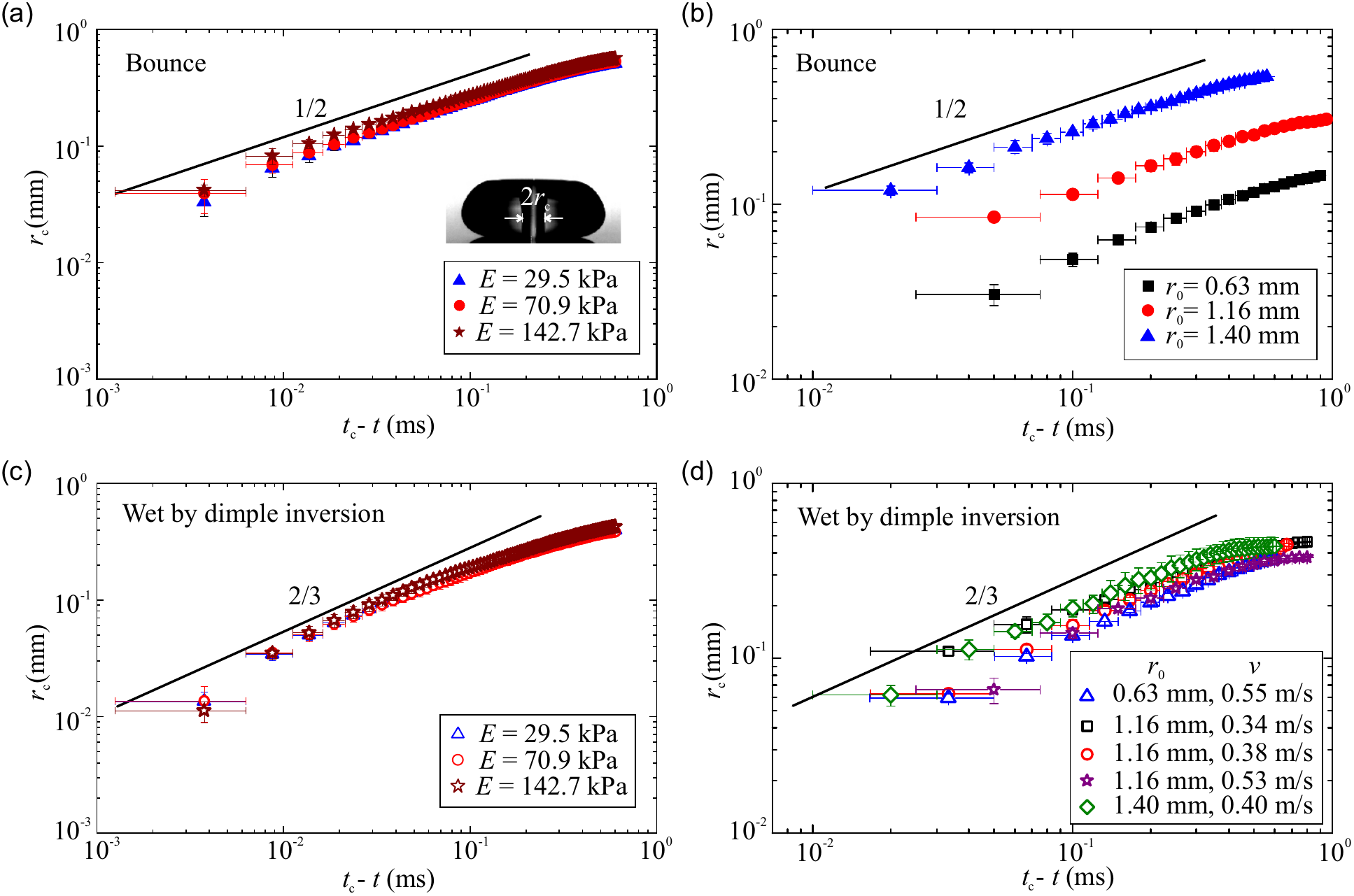}
\end{center}
\caption{(Color online) 
Plots showing the 
dependence of air cavity radius $r_{\rm c}$
with $t_{\rm c} - t$
for different impact outcomes, i.e., bounce or wet, and different experimental parameters: 
(a) droplets bounce, $r_0 = 1.4\,$mm, 
$v = 0.36\,{\rm m\,s^{-1}}$, 
varying $E$; 
(b) droplets bounce, 
$E = 4.8\,$kPa,
${\rm We} \approx 2$,
varying $r_0$;
(c) droplets wet the solid by dimple inversion, 
$r_0 = 1.4\,$mm, 
$v = 0.39\,{\rm m\,s^{-1}}$,
varying $E$;
(d) droplet wet the solid by dimple inversion,
$E = 4.8\,$kPa,
varying both $r_0$ 
and impact velocity $v$.}
\label{fig:5}
\end{figure*}

To quantitatively characterize
the collapsing dynamics of the air cavity,
we measure the
evolution of the 
retracting cavity radius $r_{\rm c}$.
In Fig.~\ref{fig:5}, 
we show the dependence of
$r_{\rm c}$ on $t_{\rm c} - t$ 
for different impact outcomes, 
i.e., bounce or wet, 
while varying the
surface elasticity $E$,
droplet radius $r_0$, 
and impact velocity $v$.
Here, $r_{\rm c}$ is the cavity radius 
measured 
along the location where
the air cavity eventually collapses, 
$t_{\rm c}$ 
the time instant 
when the retracting 
cavity completely collapses.
It should be noted here that
we only show values 
of $r_{\rm c}$ measured in the case
that the air cavity symmetrically collapses,  
which case occurs when the droplets
bounces or wet the substrate by dimple inversion
(Fig.~\ref{fig:4}b-e).
In the case that 
random rupture occurs, 
it is not possible to accurately
measure $r_{\rm c}$ 
as the cavity collapses asymmetrically
(Fig.~\ref{fig:4}f).
We observe
two distinct collapsing behaviours 
depending on the impact outcomes:
$r_{\rm c} \sim (t_{\rm c} - t)^{1/2}$ 
for bouncing droplets
(Fig.~\ref{fig:5}a,b),
and $r_{\rm c} \sim (t_{\rm c} - t)^{2/3}$ 
for wetting induced
by dimple inversion (Fig.~\ref{fig:5}c,d).
We observe that elasticity of the 
substrate $E$ does not affect the 
dependence of 
$r_{\rm c}$ on $t_{\rm c} - t$ 
in both bouncing and 
dimple inversion regimes.
In other words,
although surface elasticity significantly 
affects the transition between bouncing and dimple inversion, 
the formation and collapse 
of the air cavity is mainly determined 
by the hydrodynamical properties of the droplets.

For impacts in the bouncing regime (Figs.~\ref{fig:2}a and ~\ref{fig:4}a,b), 
it was shown 
in previous studies \cite{bartolo2006singular,chen2017submillimeter,plesset1977bubble}
that the air cavity's dynamics
is dominated by inertia,
leading to the relation 
$r_{\rm c} \sim (\gamma r_{0})^{1/4}{\rho}^{-1/4} (t_{\rm c} - t)^{1/2}$.
This is indeed consistent with 
the power law 
$r_{c} \sim A  (t_{\rm c} - t)^{1/2}$
observed in our experiment 
(Fig.~\ref{fig:5}a,b).
Fitting this power law 
to our experimental data
collected for $r_{0}$ in the range from $0.63\,\rm{mm}$
to $1.4\,\rm{mm}$ respectively yields
$A$ from $0.010$ to $0.022$
consistent with the calculated values of 
$(\gamma r_{0})^{1/4}{\rho}^{-1/4}$  
from $0.014$ to $0.020$.

For impacts in which the air film
ruptures due to dimple inversion,
we observe a different power-law behaviour
for the collapsing cavity radius.
We argue that 
the fast contact line motion 
at the rupture point
generates capillary waves
on the cavity's surface (Fig.~\ref{fig:4}); 
these waves act as the driving factor 
for the subsequent collapse of the cavity.
As the collapsing dynamics is 
in the inertial regime,
the capillary waves 
are self-similar 
with phase velocity 
$u \sim \gamma^{1/3}[\rho (t_{\rm c} - t)]^{-1/3}$ 
 \cite{day1998self,leppinen2003capillary,cox1986dynamics,vo2021dynamics}
 in the present context.
Consequently, balancing the
dynamical pressure $\rho u^{2}$
with the Laplace pressure $\gamma r_{\rm c}^{-1}$ 
at the collapsing position
results in an expression for retracting cavity radius:
\begin{equation}
r_{\rm c} \sim {{\left(\frac{\gamma}{\rho}\right)}^{1/3}}{(t_{\rm c} - t)^{2/3}}.
\label{eqn:1}
\end{equation}  
Eq.~\ref{eqn:1} indicates that 
the dependence of $r_{\rm c}$ 
on $t_{\rm c} - t$ 
in the case that dimple inversion happens
does not depend on either droplet radius
$r_0$ or the impact velocity $v$,
consistent with our experimental data
shown in Fig.~\ref{fig:5}d.

\subsection{Liquid jet dynamics}
\begin{figure*}[t]
  \begin{center}
    \includegraphics[width=0.65\textwidth]{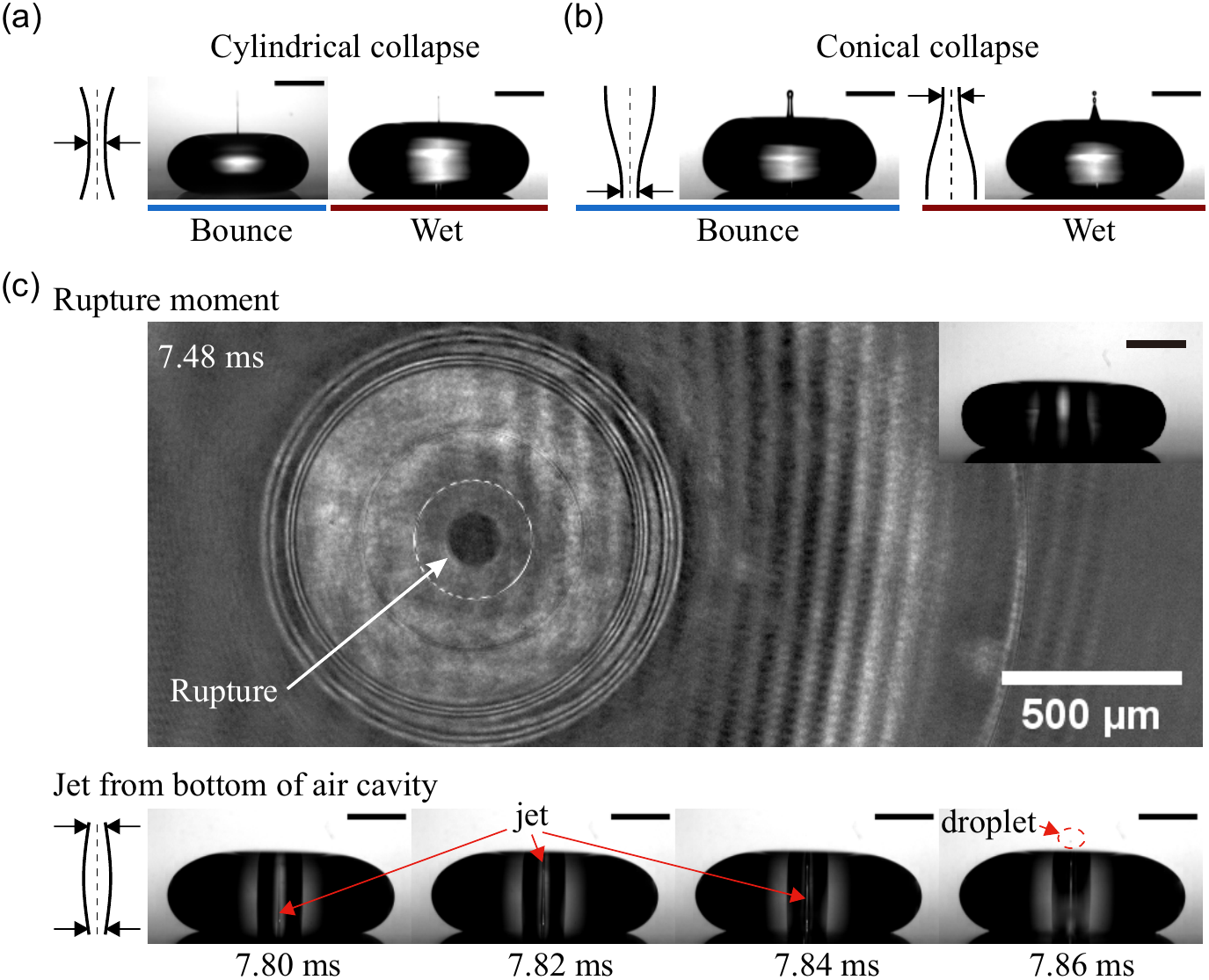}
    \end{center}
\caption{(a) Snapshots showing 
the jetting dynamics by air cavity collapsing
at its waist.
Parameters: 
for the left snapshot (bounce),
$E = 4.8\,$kPa, $v = 0.34\,{\rm m\,s^{-1}}$, $r_0 = 1.16\,$mm;
for the right snapshot (wet), $E = 70.9\,$kPa, $v = 0.38\,{\rm m\,s^{-1}}$, $r_0 = 1.4\,$mm.
(b) Snapshots showing 
the jetting dynamics by air cavity collapsing
at its ends, i.e., bottom end, top end or both.
Parameters: 
for the left snapshot (bounce), $E = 70.9\,$kPa, 
$v = 0.36\,{\rm m\,s^{-1}}$, $r_0 = 1.4\,$mm; 
for the right snapshot (wet)
$E = 29.5\,$kPa, $v = 0.45\,{\rm m\,s^{-1}}$, $r_0 = 1.4\,$mm;
(c) Snapshots showing 
the rupture moment 
by dimple inversion (upper panel) 
and the subsequent high-speed jet (lower panel) of a special case in which 
the jet originates from 
the bottom of the air cavity.
This suggests that the 
high-speed jet originating from 
the bottom of the air cavity
directly relates to 
dimple inversion phenomenon.
At the time 
the jet appears in the air cavity, i.e., at $t = 7.8\,$ms, 
the liquid already wets 
all the impact area 
(snapshots not shown here).
Parameters:  $E = 4.8\,$kPa, $v = 0.47\,{\rm m\,s^{-1}}$, $r_0 = 1.4\,$mm.
The scale bars in all the side-view snapshots represent 1\,mm.}
\label{fig:6}
\end{figure*}

When the air cavity collapses due to necking,
it is divided
into two air cavities
which subsequently evolve 
differently under 
surface tension effects.
On one hand, the enclosed geometry of 
the lower air cavity 
prevents the air from 
escaping and
eventually results
in an air bubble entrapped
inside the liquid bulk.
On the other hand, 
the upper air cavity,
which is connected 
to the ambient air at
its top continues to collapse
and finally generates 
a liquid jet 
as illustrated
in Fig.~\ref{fig:1}b.
As a result, the jetting dynamics, 
i.e., the jet's radius and velocity,
strongly depends on 
the collapsing dynamics 
of the air cavity.

In Fig.~\ref{fig:6}a and b,
we show two different 
jetting dynamics 
as a result of air cavity 
collapse dynamics, i.e., cylindrical cavity 
(Fig.~\ref{fig:4}c) 
or conical cavity (Fig.~\ref{fig:4}b, d).
When jetting is induced by 
cylindrical air cavity collapse, 
we observe that  
the jets have
 small radius and 
high velocity (Fig.~\ref{fig:6}a).
The jet's shape does not 
depend 
on whether the droplets 
bounce or wet the solid.
However, we observe that
the jetting velocity
is much higher when the
droplets wet the solid 
by dimple inversion
compared to that when it bounces.
For wetting by dimple inversion, 
we observe jet velocity
as high as $37\,{\rm m\,s^{-1}}$. 
Whereas, for bouncing scenarios,
the jet velocity observed is
typically in the range $2\,{\rm m\,s^{-1}}-10\,{\rm m\,s^{-1}}$.
When the air cavity
collapses at its ends, the so-called
conical collapse (Fig.~\ref{fig:6}b),
the jets have larger radius
and lower velocity 
compared to those 
of the cylindrical
collapse.
Typical jet velocity for conical
collapse is approximately 2\,${\rm m\,s^{-1}}$.
Whether the air film ruptures or not
does not significantly
affect either the radius 
or velocity of the jets.
The maximum jet velocity observed
in our experiments is 
approximately 100 times the 
impact velocity.
This is higher than 
the maximum jet velocity 
of droplets impact on hydrophilic, 
hydrophobic,
or superhydrophobic surfaces
reported in previous studies 
\cite{bartolo2006singular,
chen2017submillimeter,guo2020oblique,
siddique2020jet,yamamoto2018initiation},
i.e., $16 - 55$ times 
the impact velocity.
We note that 
the lower bound of jet velocity
in our case, i.e., approximately 1\,${\rm m.s^{-1}}$ 
is similar to those 
observed in the mentioned studies
\cite{bartolo2006singular,
chen2017submillimeter,guo2020oblique,
siddique2020jet,yamamoto2018initiation}.
The jet velocity appears 
to be more dependent on
the nature of air cavity at collapse, i.e., conical or cylindrical,
rather than the elasticity of the soft surfaces or their hydrophobic character.

For the lower air cavity which fails 
to escape, we consistently observe
that it eventually gets trapped
inside the liquid bulk as an air bubble.
From our experiments, we find the radius of the 
entrapped air bubble
to vary between 100\,$\mu$m - 400\,$\mu$m.
The entrapped air bubble
freely floats around in the liquid
bulk.

In Fig.~\ref{fig:6}c,
we show snapshots 
of a special case 
in which 
a thin liquid jet 
emanates from 
the bottom of the retracting
air cavity
and shoots upwards ($t \ge 7.8\,$ms)
after wetting initiates
at the center due to dimple inversion 
($t = 7.48\,$ms).
This jet is similar to the one
reported by Chen et~al. \cite{chen2017submillimeter} for impacting of droplets on structured PDMS surfaces.
We note here that the onset of
the jet occurs approximately 
$0.30 - 0.35\,$ms
after air film rupture and
wetting initiation 
suggesting that
the jet directly links
to dimple inversion causing 
air film rupture 
at the impact center.
Typically, the jet radius 
varies between 15\,$\upmu$m and 45\,$\upmu$m.
It appears that
the liquid jet expands radially briefly,
for a duration of 
60\,$\upmu$s - 80\,$\upmu$s,
before terminating
into tiny droplets.
The liquid jetting is immediately followed
by air cavity collapse and its
accompanying jet. 
Hence, for such impact scenarios,
we observe jets from
both thin liquid film collapse
and air cavity collapse.
 
\section{Concluding remarks}
In summary, we have experimentally investigated 
the impact characteristics of water droplets
on soft PDMS surfaces at low velocity.
We observed that soft surfaces promotes
bouncing for a higher range of impact Weber
number as compared to smooth hydrophilic
or hydrophobic surfaces \cite{de2015air,de2015wettability}.
Further, using interference, we probed
the interaction between the 
collapsing air cavity in the bulk of 
an impacting droplet and the air film 
separating the droplet
with the impacted soft surface. 
Above a threshold impact velocity,
the interaction induces inversion 
of the dimple in the air film 
and subsequently 
ruptures the air film. 
This behaviour is in 
a stark contrast to 
wetting initiation caused by
random air film 
rupture commonly witnessed  \cite{de2015air,langley2020droplet}. 
We then constructed a bouncing-to-wetting phase diagram 
based on the observed air film rupture dynamics. 
We experimentally confirmed 
that for impact of water droplets 
on soft surfaces, 
bouncing-to-wetting transition 
is determined 
by interactions between
the air cavity in the bulk 
and the air film trapped 
beneath the droplets, 
the so-called dimple inversion phenomenon.
The stiffness of the 
substrates, 
on the other hand, does not 
significantly alter 
the bouncing-to-wetting transition.
We have also found that, 
rupture of the air film 
by dimple inversion, 
in-turn, pins the base 
of the retracting bulk cavity 
to the surface and 
influences its collapsing dynamics. 
In such scenarios, 
the collapsing radius 
exhibits self-similarity and obeys
a $2/3$ power-law in time,
similar to 
the inertia-capillary 
break-away of a 
tapered fluid sheet \cite{keller1983surface}
or other self-similar phenomena observed
in free surface flows \cite{day1998self,lister1998capillary}.
For such collapsing mechanism,
we further observed that the 
lower half of the collapsed air cavity 
causes entrapment and formation of 
an air bubble, while 
the upper half induces 
a liquid jet with 
velocity which can be as high as
100 folds of the impact velocity.

We highlight that within the tested velocity range, 
dimple inversion and its subsequent rupture 
causes wetting initiation at the 
center, as opposed to 
random rupture locations \cite{de2015air,langley2020droplet}. 
This mode of liquid-solid contact is a more 
controlled event where the liquid-solid 
footprint radius 
expands in a radially symmetric manner, 
similar to onset of quasi-static spreading of 
liquid drops on surfaces \cite{biance2004first,mitra2016understanding}.
Such controlled wetting characteristics 
may contribute to
minimizing air bubble entrapment
upon droplet impact 
and thus would be helpful
in applications such as droplet deposition \cite{breitenbach2018drop} 
or ink-jet printing
\cite{ashgriz2011handbook}.


\section*{Conflicts of interest}
There are no conflicts to declare.

\section*{Acknowledgments}
The authors thank Maurice H. W. Hendrix for valuable discussions regarding the procedure to extract dimple profile shape from interference fringes. 
The authors also thank Marcus Lin for his help with rheometer measurements of the soft surfaces.
This study is supported by Nanyang Technological University (NTU) and A*STAR (SERC Grant No. 1523700102). S.M. is supported by NTU Research Scholarship.

\providecommand*{\mcitethebibliography}{\thebibliography}
\csname @ifundefined\endcsname{endmcitethebibliography}
{\let\endmcitethebibliography\endthebibliography}{}

\end{document}